\documentstyle[twoside,fleqn,espcrc2,rotate,epsf,array]{article}

\def\bsp{{{\mbox{\boldmath${\sigma}$}}}}
\def\bmp{{{\mbox{\boldmath${\mu}$}}}}
\def\bvp{{\hat {\mbox{\boldmath${p}$}}}}
\def\lsim{{\buildrel < \over \sim}}

\title{Transport through superconductor/magnetic dot/superconductor structures}

\author{M.~Andersson\address[C]{Institute of Theoretical Physics, 
                                Chalmers University of Technology
                                and G\"oteborg University,
                                S-412 96 Gothenburg, Sweden},
        J. C. Cuevas\address[K]{Institut f\"ur Theoretische Festk\"orperphysik, 
                                Universit\"at Karlsruhe, D-76128 Karlsruhe, Germany},
        and
        M. Fogelstr\"om\addressmark[C]
       }
\begin{document}

\begin{abstract}
The coupling of two s-wave superconductors through a small magnetic dot is discussed.
Assuming that the dot charging energy is small compared to the superconducting
gap, $E_c\ll \Delta$, and that the moment of the dot is classical, we develop a simple theory
of transport through the dot. The presence of the magnetic dot will position Andreev bound 
states within the superconducting gap at energies tunable with the
magnetic properties of the dot.
Studying the Josephson coupling it is shown that the constructed junction can be
tuned from a "$0$" to a "$\pi$"-junction via a degenerate two-level state either by
changing the magnetic moment of the dot or by changing temperature. Furthermore, it
is shown that details of the magnetic dot can be extracted from
the sub-harmonic structure in the current-voltage characteristics of the junction.
\end{abstract}

\maketitle
 
If a superconductor is exposed to magnetically active impurities \cite{shiba,BKS77} 
or materials \cite{BBP82} the
superconducting state is modified. Coupling two superconductors through
a magnetically active barrier may lead to what is known as a $"\pi"$-junction \cite{BKS77},
a junction for which the ground state has an internal phase shift of $\pi$ between the
superconductors across the barrier.
If the barrier is extended to an S/F/S-structure, a ferromagnetic (F) layer sandwiched between
two superconductors (S), the critical current will oscillate as the thickness of F is varied \cite{BBP82}.
Recently, this behavior was reported in a mesoscopic S/F/S-structure and
a systematic change from a "0" to "$\pi$" Josephson junction as a function of 
the F-layer thickness was seen \cite{Ryazanov01}. In this
paper we present a different scenario which gives rise to qualitatively the same physics, namely,
two s-wave superconductors connected through a small magnetic dot. In this case the effect of the
magnetic dot arises from the spin-active contact at the interface between the dot and the
superconductor in contrast to the internal exchange field of the ferromagnet as in Refs. 
\cite{BBP82,Ryazanov01}.

\begin{figure}[htb]
\hspace{9pt}
\epsfxsize=0.43\textwidth{\epsfbox{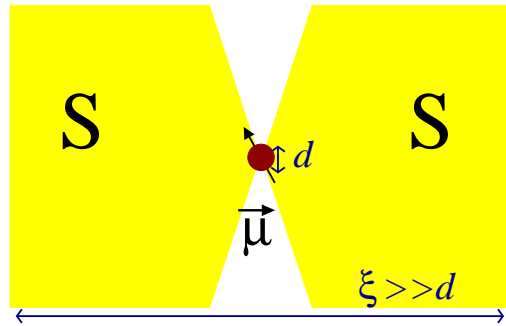}}
\vspace*{-20pt}
\caption{A schematic illustration of the superconductor/magnetic dot/superconductor
contact considered. The size of the dot, $d$ is small compared to the superconducting
coherence length $\xi_0$. It is also assumed that the moment, $\vec{\bmp}$, is large, i.e. in the
classical limit.}
\vspace*{-13pt}
\label{fig:pinhole}
\end{figure}
In Figure \ref{fig:pinhole} we show a cartoon of our system:
two {\em conventional} superconductors brought in electrical contact through a magnetic
grain. The size of the grain, $d$, is small compared to the superconducting coherence length, $\xi_0$,
so that  $d/\xi_0\ll 1$ holds.
The dot is characterized by a large classical magnetic moment, $\vec{\bmp}$, and
despite the small size of the grain we shall assume that
charging effects on the dot can be neglected, i.e. $E_C\ll\Delta$.
In our approach the effect of the magnetic dot enters via boundary conditions for the quasiclassical
Green's function. In this spirit the
contact over the magnetic dot is described by the following $\hat S$-matrix 
\begin{equation}
\hat S=
\left(\begin{array}{cr}r&t\\t&-r\end{array}\right)\exp(i\frac{\Theta}{2}{\hat \bsp}_{\!\bmp})
\label{eq:smatrix}
\end{equation}
were $r$ and $t$ are the usual reflection and transmission coefficients. The spin-dependence
enters via a rotation of the quasiparticle spin by an angle $\Theta$ around the moment axis,
${\hat \bsp}_{\!\bmp}$. 
The angle $\Theta$ is a phenomenological parameter that quantifies the degree of "spin-mixing"
induced by the moment $\vec{\bmp}$ and
varies between $0$ (no spin-mixing) and $\pi$ (strong spin-mixing) \cite{Tokuyasu88,Fogel00}.
The calculations presented here are done within the quasiclassical theory as presented in
Refs. \cite{Fogel00,Cuevas01} and references therein.

\paragraph{Local quasiparticle density of states.}
\noindent
One of the most pronounced effects of the moment, $\vec{\bmp}$, on the superconducting density of states
(DOS) is the formation of Andreev states 
at $\varepsilon_B(\Theta)=\pm \Delta \cos(\Theta/2)$.
The two signs refer to the two "spin bands" on which the pairing amplitudes are
$\sim \langle \psi_\uparrow \psi_\downarrow\rangle$ $(+)$
and $\sim\langle \psi_\downarrow \psi_\uparrow\rangle$ $(-)$ respectively.
The formation of these Andreev states is quite easily understood since the
moment will give reflected quasiparticles 
a phase shift of $\exp{(\pm i \Theta)}$. The sign of the phase shift depends on
the spin of the quasiparticle: $(+)$  for spin up and $(-)$ for spin down.
Introducing the moment axis, ${\hat \bsp}_{\!\bmp}$,
breaks the rotational symmetry in spin-space of the superconductors and pins down 
the direction of the spin-quantization axis to be parallel to ${\hat \bsp}_{\!\bmp}$.
 
Formation of Andreev states is a signature of a reduced superconducting order by pair-breaking.
The Andreev states are spatially localized at the magnetic moment
and their spectral weight decays exponentially in to the superconductors over a distance given by  
$\xi(\varepsilon_B)=v_F/2 \sqrt{\Delta^2\!-\!\varepsilon_B^2}$, where $v_F$ is the Fermi velocity 
in the superconductor. This gives that superconductivity heals to its bulk value over a distance 
$\sim \xi(\varepsilon_B)$
into the superconductor.
However, in the limit $d/\xi_0\ll 1$ the pair-breaking
effect of the dot on the superconducting order parameter is small
and we can, to good approximation,  assume a constant
order parameter up to the contact. Another consequence of the relative smallness $(d/\xi_0\ll1)$
of the dot is that the broadening of the Andreev state by elastic impurity scattering is suppressed.

\begin{figure}[htb]
\hspace{-3pt}
\rotate[r]{\epsfxsize=0.33\textwidth{\epsfbox{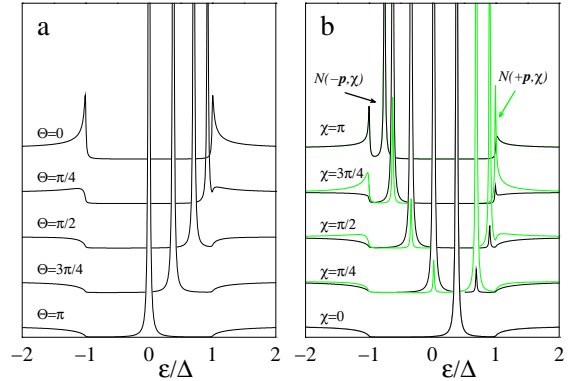}}}
\vspace*{-14pt}
\caption{The density of states (DOS) at a magnetic dot for different values of $\Theta$ in (a). The
bound states appear at $\varepsilon_B=\pm\Delta \cos (\Theta/2)$, the different signs
refer to two the spin bands (see text). Here we plot $\varepsilon_B=+\Delta \cos (\Theta/2)$.
In panel (b) we display the DOS for the momenta $\pm\bvp$, on the same spin-band as in (a),
with $\Theta=3\pi/4$ for various values of phase difference $\chi$ over the junction, 
${\cal{D}}=0.9$.} 
\vspace*{-13pt}
\label{fig:SpinDos}
\end{figure}

It is important to note that the positions of the bound states
do not disperse with momentum ${\bvp}$ but do only depend on the 
angle  $\Theta$. This makes the position of states $\varepsilon_B$ robust
and the DOS shown in Figure \ref{fig:SpinDos}.a are angle averaged
ones at the dot. Also important to note is that the DOS in both
superconductors are symmetric and $\varepsilon_B$ coincide in both
sides. This makes the DOS independent of the transmission probability,
${\cal{D}}$, as long as there is no phase difference $\chi$ applied across
the junction.

\paragraph{Tunable Josephson-junction properties.}
\noindent
It is well known that the Josephson current is carried by junction Andreev states
\cite{swave,Riedel98,Barash00}. In case of a magnetic dot the junction Andreev states
split into four states, labelled by their momentum directions $\pm\bvp$ and by spin-band,
as a phase difference $\chi$ is applied across the junction \cite{Cuevas01}. 
We have 
\begin{equation}
\begin{array}{rl}
\varepsilon_B(\chi) =
\pm& \!\!\!\!\!\Delta \bigg[  
\cos^2(\frac{\Theta}{2}) -{\cal {D}} \cos(\Theta) \sin^2(\frac{\chi}{2})\\
\pm\sqrt{{\cal {D}}}&\!\!\!\!\!\sin(\Theta) \sin(\frac{\chi}{2}) 
\sqrt{1 - {\cal {D}} \sin^2(\frac{\chi}{2})}\bigg]^{\frac{1}{2}}
\end{array}
\label{eq:phasebs}
\end{equation}
which now not only depends on $\Theta$ but also on $\chi$ and  ${\cal{D}}$.
In Figure \ref{fig:SpinDos}.b we show the DOS for the two opposite
momentum directions, $+\bvp$ and $-\bvp$, on the positive 
spin-band with $\Theta=3\pi/4$, for various $\chi$ at ${\cal{D}}=0.9$. 
As seen in Figure \ref{fig:SpinDos}.b the Andreev states in the DOS splits 
equally for {\em both} $+\bvp$ and $-\bvp$ with applied phase difference $\chi$
as given by Eq.(\ref{eq:phasebs}). The spectral weight of the bound states,
$w(\varepsilon_B(\chi,\bvp))$, however, is not symmetric in momenta,
i.e. $w(\varepsilon_B(\chi,-\bvp))\ne w(\varepsilon_B(\chi,+\bvp))$. This leads
to, after summing up all contributions,
a net Josephson current that may have additional zeros in the phase interval 
$0$ and $\pi$.

The sensitivity of the quasiparticle spectra in Eq.(\ref{eq:phasebs}) to the degree 
of spin-mixing, $\Theta$, and to the transmission probability 
${\cal{D}}$ at finite phase difference, makes
the magnetic dot a very tunable Josephson contact between the two superconductors it connects.
In d-wave superconductors, as the high-$T_c$ cuprates,
Andreev states at zero-energy lead to anomalous Josephson properties \cite{dwaveanomT,dwaveanomP}.
In an ideal junction 
between two d-wave superconductors the zero-energy states will cause
the critical current to diverge $\sim 1/T$ at low temperature \cite{dwaveanomT}.
In addition, the energetics of the junction may change from being a "0"-junction at temperatures
just below $T_c$ to a "$\pi$"-junction at low temperatures $T \ll T_c$ \cite{dwaveanomT,dwaveanomP}. 

\begin{figure}[htb]
\centerline{\epsfxsize=0.45\textwidth{\epsfbox{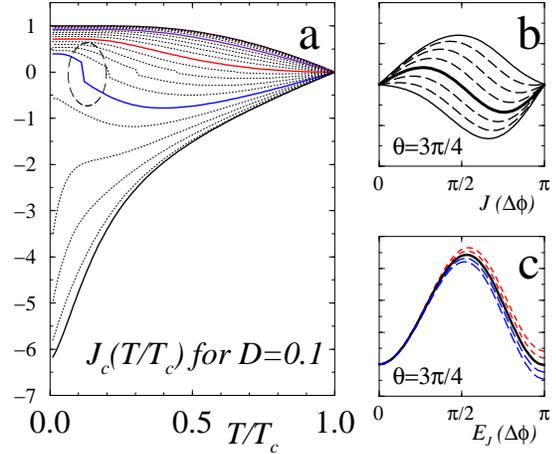}}}
\vspace*{-25pt}
\caption{The critical current vs temperature of a magnetic-dot Josephson contact is shown in (a). The
spin-mixing angle, $\Theta$ is varied from $0$ to $\pi$ (top to bottom) in steps of $\pi/20$. 
The transmission probability
${\cal{D}}$ is $0.1$ (${\cal{D}}=|t|^2$). In panels (b) and (c) we show the current-phase 
relation (b) and the
junction energy vs. phase (c) for $\Theta=3\pi/4$ at temperatures just around the switching temperature 
$T_{sw}\sim 0.12 T_c$, highlighted by the dashed circle. At $T_{sw}$ the junction changes state from a 
"$\pi$" to a "0"-junction. As seen, the junction goes through a degenerate $\pi$-periodic state at the
switching temperature.
}
\vspace*{-12pt}
\label{fig:Josephson}
\end{figure}
 
As demonstrated in Refs. \cite{Fogel00,Cuevas01} the Josephson coupling of two 
s-wave superconductors through a 
magnetically active barrier shows much of the same anomalies as seen in the d-wave/d-wave contact.
In Figure \ref{fig:Josephson}.a we show the critical current as a function of temperature
for several values of $\Theta$ ranging for 0 to $\pi$. At low temperatures
and for $\Theta$ close to $\pi$ the critical current is roughly an order of magnitude larger
than the value of a non-magnetic dot, i.e. at $\Theta=0$. In fact, it is a finite transparency,
in Figure \ref{fig:Josephson}.a
put to ${\cal{D}}=0.1$, or an inelastic or phase-breaking life time, $\tau_{inel,pb}$, 
depending on which 
gives the largest broadening of the junction states, that cuts of the growing critical current 
as $T\rightarrow0$ \cite{Fogel00,Cuevas01}.  

In Figures \ref{fig:Josephson}.b-c we show the current-phase and the energy-phase relations of a 
junction with $\Theta=3\pi/4$ and ${\cal{D}}=0.1$. The temperature is swept over a small interval
including $T_{sw}$. At $T=T_{sw}$ the junction changes its energy state from being a "0"-junction
at low temperatures to being a "$\pi$"-junction at $T>T_{sw}$. This transition goes over
$\pi$-periodic current-phase relation at $T=T_{sw}$ and here the "0" and "$\pi$" states of the junction
are degenerate. A similar transition may be achieved at fixed $T$ and instead sweeping $\Theta$.

\begin{figure}[thb]
\hspace{5pt}
\rotate[r]{\epsfxsize=0.37\textwidth{\epsfbox{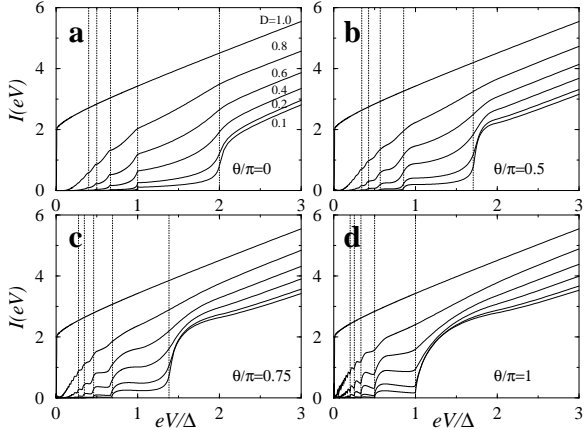}}}
\vspace*{-30pt}
\caption{The current-voltage characteristics for the magnetic Josephson contact shown for 
$\Theta=0, \pi/2 ,3 \pi/4,$ and $\Theta=\pi$ in panels (a) through (d). In each panel we show
the IV-curve for different transparencies as labeled in panel (a). The IV-curves in panel (a)
correspond to an ordinary single-channel point contact between two s-wave superconductors. The
appearance of Andreev states at $\varepsilon_B=\pm \Delta \cos (\Theta/2)$ changes the sub-gap structure 
of the magnetic dot (b-d) as described in Eq.(\ref{eq:iv}) and here marked by the vertical dashed lines.
In each case the current is given in units $(G_N\Delta/e)$ where $G_N$ is the normal conductance.
}
\vspace*{-12pt}
\label{fig:IV_vs_Theta}
\end{figure}

\paragraph{Current-voltage characteristics.}
\noindent
As introduced, the spin-mixing angle is a phenomenological parameter in the
same spirit as the junction transmission probability is. In principle one should be able
to give a microscopic justification of the value of $\Theta$
by calculation along the lines in Ref. \cite{Tokuyasu88}. In a practical realization 
of a magnetic dot it is more likely that an effective $\Theta$ is measured and it is important
when characterizing the junction
to extract the value of $\Theta$ by comparing measurements with calculations. 
Pinning down $\Theta$ and ${\cal{D}}$ can be done by measuring several junction specific
properties such as the Josephson current, the current-voltage characteristics and the noise
spectrum of the junction as was recently done in Al point contacts \cite{Cron01}. Here
we compute one of these properties, the dc-current-voltage curves, for a single
channel contact as a function of transparency, ${\cal{D}}$, and spin-mixing, $\Theta$, using
the methods described in Ref. \cite{Cuevas01}.

At a constant voltage bias, $|e V|<2 \Delta$, applied over a junction between two superconductors the only
way a dc-current can flow through the contact is via multiple Andreev reflection (MAR) 
\cite{Klapwijk82}.
This simply means that in order to transfer a quasiparticle at energy
$\varepsilon =-\Delta$ in the superconductor at bias $-|e V|/2$ to  
energy $\varepsilon = \Delta$ in the superconductor at bias $+|e V|/2$ 
the quasiparticle must undergo $2 \Delta /|e V|$ Andreev
reflections. At each reflection event the quasiparticle is accelerated and shifted up
in energy by an amount $|e V|$. In the current-voltage characteristics this shows up in
an abrupt increase in the current each time the voltage is swept over a value
where $\mbox{mod} (2 \Delta ,|e V|)$ changes. This holds true for the conventional superconductor
for which there are no states within the superconducting gap \cite{swaveiv,atomic}. The resulting
sub-gap structure in the current-voltage curve
for $\Theta=0$ is shown in Figure \ref{fig:IV_vs_Theta}.a for several values of ${\cal{D}}$.

Introducing Andreev states within the gap, by a 
finite value of $\Theta$, we modify the 
MAR-picture above. Now, having 
the bound state $\varepsilon_B(\Theta)=\pm \Delta \cos(\Theta/2)$ and 
starting at a large voltage bias $|e V| \lsim 2 \Delta$,
the first voltage 
below $2 \Delta$ to connect two points in the quasiparticle spectra,
on either side of the junction, with a finite DOS, and
thus to give a contribution to the current, is at
$|e V|_1=(1 +\cos(\Theta/2)) \Delta $. 
There are two possible processes giving current at $|e V|_1$ per spin-band. For the
band with $\varepsilon_B(\Theta)=+ \Delta \cos(\Theta/2)$ we have: (i) connecting an
energy at $-\Delta $ with $\varepsilon_B(\Theta)$, i.e. an electron-like quasiparticle 
tunneling into the bound state
from the continuum states below $-\Delta$. (ii) connecting $\varepsilon_B(\Theta)$ with
the continuum states at $-\Delta$, i.e a hole-like quasiparticle  
tunneling from the bound state
into the continuum states below $-\Delta$.
For the spin-band with $\varepsilon_B(\Theta)=- \Delta \cos(\Theta/2)$ the same processes
occur with the modification that the quasiparticle(hole) start in the bound state and ends up
in the continuum at $+\Delta$.
The next "spectral current channel" to open 
is at voltages $|e V|_2=(1 +\cos(\Theta/2)) \frac{\Delta}{2}$ which, via a single Andreev reflection
off the opposing superconductor,
connects finite DOS in the same electrode. Adding more and more spectral current channels,
$|e V|_n$, we arrive at 
\begin{equation}
|e V|_n =\bigg(1+\cos(\frac{\Theta}{2})\bigg)\,\frac{\Delta}{n}
\label{eq:iv}
\end{equation}
which gives the location of the strong features in the current-voltage curves at different 
spin-mixing angles, $\Theta$. This is clearly demonstrated in Figure \ref{fig:IV_vs_Theta} where
the position of the sub-gap structures described by Eq.(\ref{eq:iv}) are marked.

Why do you not see structure in
$|eV|_n = 2\Delta/n$ as well? In principle, one could think that the usual 
subharmonic gap structure should be superimposed on this new one. 
The answer is that you still have the usual processes between $-\Delta$ and 
$+\Delta$, but now the DOS at the gap edges is rather small. Thus,
opening the possibility for one of these processes, its probability
is small (although its contribution will increase with voltage).

Finally we should mention that the IV-curves in Figure \ref{fig:IV_vs_Theta} are calculated 
with a very small intrinsic broadening, $1/2 \tau_{inel,pb}=10^{-4} \Delta$. Increasing 
$1/2 \tau_{inel,pb}$ smears out the sharp features at low voltages in to a large current peak.
This is particularly strong when $\varepsilon_B\approx 0$, i.e. for $\Theta$ close to $\pi$.
At $\Theta=\pi$
the IV-curves are similar to those of a symmetric, 45-degree rotated, d-wave/d-wave junction 
\cite{Lofwander99,Cuevas01}. 

\paragraph{Conclusions and acknowledgements.}
In conclusion we have presented a simple theory to handle transport through magnetic dots
connecting two conventional superconductors in the limit where charging effects may 
be neglected. We show that the Andreev bound state spectra is tunable with the properties
of the dot.  This allows realizations of Josephson junctions that are either "$0$" or 
"$\pi$"-junctions.
Furthermore, we look at ways to extract information of the dot from
experimental data and in detail study the current-voltage characteristics. The 
presence of Andreev
states in the DOS allows for current through resonant tunneling \cite{Resonant} and modifies
the sub-gap structure in the IV-curves compared to the case of a conventional SIS-contact.

We are happy to acknowledge enlightening discussions with Tomas L\"ofwander.
The research was founded by the Swedish Research Council (VR) (M.~A. and M.~F.) and by 
the DFG project SFB 195 and the EU LSF programme (J.~C.~C.).

\end{document}